# Interplay between structure and relaxation in polyurea networks: the point of view from a novel method of cooperativity analysis of dielectric response

Nerea Sebastián,[a,b] Christophe Contal,[c] Antoni Sánchez-Ferrer,[*,d] Marco Pieruccini[*,e]

The influence of structural constraints on the relaxation dynamics of three polyurea networks with varying degree of crosslinking, has been studied by means of a thorough analysis of broadband dielectric spectroscopy measurements. Two different relaxation processes are observed, namely, a fast process involving the soft poly(propylene oxide) chains, and a slower and much broader process associated to the immediate surroundings of the hard crosslinkers. Microphase separation in soft and hard domains characterizes the systems in the presence of hydrogen bondings. In this case, different confinement conditions are explored by varying the soft chain length; overall, the so called "adsorption" effects dominate. With respect to both cooperativity and rearrangement energy threshold in fast relaxation, it is found that the enhancement of configurational constraints is similar to cooling, but only on qualitative grounds. An upper bound of the hard domains' interface thickness, in which the slow relaxation is believed to take place, is estimated from the analysis of the fast relaxation in the system characterized by the highest degree of confinement, taking into account the results of the structural analysis. Dropping the hydrogen bonding mechanism, phase separation does not occur anymore and the configurational constraints at the ends of the soft chains are reduced, leaving just those imposed by the rigid crosslinkers. This leads to a significant increase in cooperativity, on approaching the glass transition, and to a complex behavior that is thoroughly discussed in comparison with those observed in the micro-segregated systems.

## 1 Introduction

Molecular dynamics in Soft Matter is a hot topic in Condensed Matter Physics; this does not only have to do with potential industrial interests, but also with knowledge in its most general and fundamental sense. Indeed, notwithstanding the considerable efforts that are being spent since many years in the field, basic aspects of this branch of Physics still remain poorly understood. Among others, the mechanisms leading to the formation of a glass from a supercooled liquid, yet represent a matter of lively debate. This situation is also a bit confusing, given that the use of glasses lasts since centuries.

When a liquid is cooled down and crystallization is avoided, its viscosity progressively increases and, sooner or later, it will form a glass: a solid-like state in which long range molecular order is absent. On approaching this state, the central relaxation time of the configurational motion often shows a super-arrhenian temperature dependence, that is, the apparent activation energy increases on cooling. This behaviour, hinted at the possibility that the configurational transitions responsible of the viscosity, might involve an ever-increasing number of rearranging units as the glassy state is approached. The quest for a characteristic length of the glass transition thus run initially through seeking its signature in structural features. Later, it was realized that this length would more appropriately relate to dynamic correlations.

Probing this length, though, is not trivial and several approaches have been (and are being) developed in order to grasp it from relaxation experiments. Thus, Donth[1,2] provided a fluctuation theory whereby a characteristic size of cooperatively rearranging regions (CRRs) in a supercooled liquid - i.e., domains where rearrangements may occur without inducing any configurational change in the rest - could be estimated from the temperature fluctuations accompanying the energy exchanges between configurational and non-configurational degrees of freedom.[3] One of the most direct evidences for the increase in CRR size on cooling, has been reported in recent investigations on Polystyrene and Poly(methyl methacrylate), based on modulated calorimetry.[4]

Another theoretical approach has been developed by Berthier and co-workers, who focused instead on the

[a.] Jožef Stefan Institute, Department of Complex Matter - F7, Jamova cesta 39, SI-1000 Ljubljana, Slovenia.
[b.] Universidad del País Vasco (UPV/EHU), Departamento de Física Applicada II, Apdo. 644, E-48080 Bilbao, Spain.
[c.] Institut Charles Sadron, Physique et Biophysique des Macromolécules aux Interfaces, 23 rue du Loess, BP 84047, F-67034 Strasbourg cedex, France.
[d.] Swiss Federal Institute of Technology, Department of Health Sciences and Technology, IFNH, Schmelzbergstrasse 9, LFO, E29 CH-8092 Zürich, Switzerland. E-mail: antoni.sanchez@hest.ethz.ch
[e.] CNR, Istituto Nanoscienze, v. Campi 213/A, I-41125 Modena, Italy. E-mail: marco.pieruccini@nano.cnr.it
† Electronic Supplementary Information (ESI) available:

generalized (4-point) susceptibility $\chi_4(\mathbf{y},t) = \langle\phi(\mathbf{r},t)\phi(\mathbf{r+y},t)\rangle$, describing correlations between relaxation dynamics, i.e. $\phi(\mathbf{x},t)$, at $\mathbf{x=r}$ and $\mathbf{x=r+y}$, or on the 3-point susceptibility $\chi_T(\mathbf{y},t) = \langle h(\mathbf{r},0)\phi(\mathbf{r+y},t)\rangle$ relating to the effect of an enthalpy fluctuation at $\mathbf{r}$ on the relaxation dynamics at $\mathbf{r+y}$[5,6] (in particular, the second one is directly accessible from the experiment, e.g., $\chi_T(\mathbf{0},t) = \partial\phi(\mathbf{0},t;T)/\partial T$. Also within this framework, evidence has been observed of a growing characteristic length on cooling a relaxing liquid.

More recently a further method has been proposed, which considers more closely the features of the energy landscape in a relaxing liquid.[3,7] In this approach, the cooperativity is derived by fitting the configurational relaxation function, extracted from the data, with a theoretical expression derived within a statistical mechanical framework. The basic idea behind is that the diffusional motion characterising the rearrangements of units over length scales in the order of the CRR size follows by facilitation from a pre-transitional dynamic state of fluctuational nature.

Also with the latter method, of course, an increase in cooperativity of a liquid (either polymeric or not) under progressive cooling towards its glass transition has been found; the agreement with Donth's estimates was good.[3,8] At the present state of the art, the last circumstance is extremely important, because in the absence of direct measurements of the cooperativity length, a figure of reliability in the analysis is represented by the comparison with the results of independent methods of estimate.

When a certain characteristic length $L_h$ is introduced in a polymeric system through appropriate structural heterogeneities (such as crosslinks, dispersed nano-particles or even the walls of confining nano-pores),[9,10] it is expected that a change in the relaxation behaviour occurs when the dynamic correlation length approaches that value. On the one hand, this circumstance provides some objective support to the existence of a cooperativity length; on the other, it allows to probe the relaxation dynamics via a proper tailoring of the system.

Elastomers are crosslinked polymer melts having characteristics from the three physicals states. They are soft solids which can be shaped due to the presence of crosslinkers which avoid the polymer melt to flow; they contain a main component which is a viscous liquid; and the retractive force appeared upon stretching is entropically driven similar to the case of gases.[11] The chain mobility in such materials decreases upon cooling down due to the thermal inactivation of conformational changes, and the reduced free volume in the surroundings of the chains.[12] Therefore, viscosity increases and the polymer chains freeze into the glassy state where no longer a conformation change is possible, but reorientation of short segments in the polymer backbone or side groups. The high viscosity value at low temperatures in elastomeric networks is the main cause of the slowing down of the dynamics of such systems, together with parameters introducing asymmetry or disorder to the system, e.g., polydispersity, crosslinking, entanglements, looping, dangling ends, supramolecular interactions.

Polyurea networks with varying degrees of crosslinking, have been recently investigated in relation to their mechanical relaxation properties.[13] In these systems, the presence of hydrogen bonding between urea linkages leads to the formation of nano-segregated hard domains, connected one with the other by flexible chains. By changing the length of the latter, different mechanical constraints to the chain's configurations were imposed. Considering systems made of soft polyetheramine chains with $M_c \sim$ 400, 2000 and 4000 g·mol$^{-1}$ segmental molecular weights, it was possible to explore structures turning from an ensemble of hard domains embedded in matrices of increasing rigidity, to a condition in which isolated domains of limited mobility are embedded in an environment of percolated hard domains (i.e., $M_c \sim$ 400 g·mol$^{-1}$). In the less constrained systems, it was possible to separate two distinct segmental relaxations: the faster one associated to the embedding matrix, and the other to a stiffer layer surrounding the hard domains (each with its own glass transition temperature). In the sample with the shortest flexible chain, these processes tended to merge into one and the glass transition temperature of the softer component increased significantly, mainly due to the more restricted mobility of the soft polymer segment when being pulled from both extremes by the interfacial polymer layer and well transmitted along the soft segmental. Such a slowing down of the dynamics can be attributed to the overwhelming effect of the interaction with the confining walls (so called "adsorption")[14], over the acceleration expected from pure confinement. This is indeed related to the soft chain pinning to the hard nano-particles.

There are several aspects which deserve being investigated in this context; one among the others is the interplay between temperature and confinement in the regime of adsorption. In other words, concerning the configurational relaxation, one may question to what extent the enhancement of the constraints can be considered similar to a decrease in temperature. Another aspect regards how things change when the constraints on soft chain ends deplete to some extent.

These two issues represent the main concern of this paper. For the first one, we shall perform dielectric relaxation analyses (necessarily in isothermal conditions), on the systems of reference [13] with $M_c \sim$ 2000 and 4000 g·mol$^{-1}$ soft chains. For the second, comparisons will be done among the relaxation dynamics thus probed, and that which establishes in a modification of the $M_c \sim$ 2000 g·mol$^{-1}$ sample in which hydrogen bondings are suppressed, thus allowing a limited mobility to the soft chain ends.

The cooperativity associated to the fast segmental process, and its impact with the structural features of the systems, will be thoroughly discussed by means the method of references [3,7]. The results will be also compared with the lower bound estimates provided by the 3-point correlation function analysis.[5,6] The discussion will be done with reference to structural information extracted by means of large and small angle X-ray scattering (WAXS and SAXS respectively) and by mappings from atomic force microscopy (AFM).

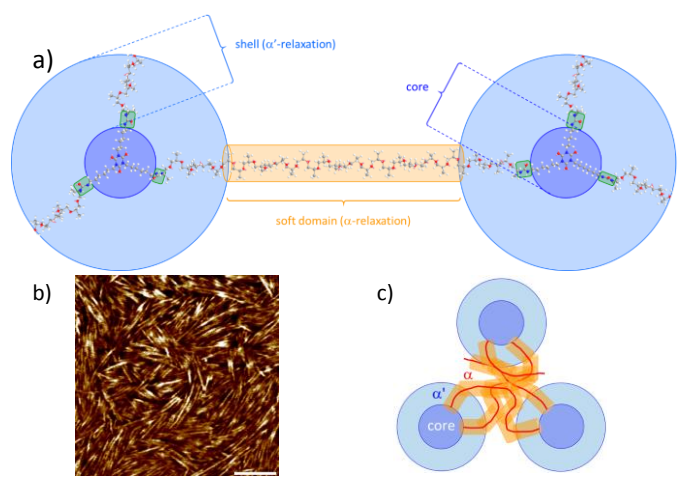

Figure 1. a) Schematic representation of the studied polyurea elastomeric system, where the soft domain (orange cylinder), the hard domain (dark blue circle), the shell (light blue corona), and the urea linkages (green) are shown. b) AFM phase profile image showing the big contrast between the hard domains (rod-like objects) embedded into the soft matrix. *Note*: scale bar is 100 nm. c) Model representation of the cross-section of the rod-like objects where the soft domain (orange), the shell (light blue) and the hard domain (dark blue) are shown.

## 2 Experimental

**Sample preparation**

The three linear hydrophobic diamino-terminated polyetheramines Jeffamine® D-2000 ($M_n$ = 2056 g mol$^{-1}$, $\rho$ = 991 kg m$^{-3}$), D-4000 ($M_n$ = 4000 g mol$^{-1}$, $\rho$ = 994 kg m$^{-3}$) and SD-2001 ($M_n$ = 2050 g mol$^{-1}$, $\rho$ = 978 kg m$^{-3}$) from Huntsman International LLC, and the triisocyanate crosslinker Basonat HI-100 ($mw$ = 504 g mol$^{-1}$, $\rho$ = 1174 kg m$^{-3}$) from BASF SE were used as received. Polyurea (PU) networks are formed by the rapid chemical reaction between a triisocyanate and a diamine to build urea moieties, as shown in Figure 1. Details of the synthesis of these PU elastomers are well described in the literature.[13,15,16] Table 1 compares the main structural characteristics for the three used samples ED-2000, ED-4000 and ESD-2001. The structure of polyurea chains is shown in Figure 1; due to strong hydrogen bonding between urea linkages of neighboring chains or neighboring segments in the same chain, the hard segments microphase segregate into so-called "hard domains". That is the case for samples ED-2000 and ED-4000, whose genitor polymer is a primary amine, but not for the sample ESD-2001 coming from the reaction between bulky secondary amine-containing polymer and the crosslinker.

**Methods**

The complex dielectric permittivity $\varepsilon^*(\omega) = \varepsilon'(\omega) - i\varepsilon''(\omega)$ was measured over a broad frequency range (10$^{-2}$-10$^9$ Hz) by combining two impedance analyzers: Alpha-A + ZG4 (Novocontrol) for low frequencies and HP4192A for high frequencies. In the latter setup (high frequencies), the cell consist of a parallel plate capacitor placed at the end of a coaxial line. Due to geometry constrains the capacitor consists of two 5 mm diameter gold plated electrodes separated by the polymer as dielectric. A modified HP16091A coaxial test fixture was used as the sample holder. In order to obtain larger capacitance values at low frequencies, larger samples of 15 mm diameter were employed for the low frequency measurements, with the sample placed between parallel plate electrodes at the end of homebuilt four-wire holder. Very careful calibration of both systems ensured well matching between the two sets of measurements. In both setups, all the three samples consisted of thin uniform films with a thickness around 0.3-0.4 mm, which were held in a cryostat. Isothermal spectra were recorded at constant temperature with a resolution of 0.1 ºC under nitrogen atmosphere.

Tapping mode atomic force microscopy (AFM) was carried out on a Multimode 8 Scanning Force Microscope (Bruker). NanoSensors AFM cantilevers for tapping mode (NCHR) in soft and hard tapping conditions were used at a typical resonant frequency of 300 kHz. Images were simply flattened and analyzed using Nanoscope Analysis 1.5 software, and no further image processing was carried out.

Simultaneous small- and wide-angle X-ray scattering (SAXS and WAXS) experiments were performed using a Philips PW 1730 rotating anode (4 kW) in order to obtain direct information on the SAXS and WAXS reflections. Cu K$_a$ radiation ($\lambda$ = 1.5418 Å) filtered by a graphite monochromator and collimated by a 0.8mm collimator was used. The incident beam was normal to the surface of the film. The scattered X-ray intensity was detected by a Schneider image plate system (700x700 pixels, 250 µm resolution). An effective scattering-vector range of 0.1 nm$^{-1}$ < $q$ < 30 nm$^{-1}$ is obtained, where $q$ is the scattering wave-vector defined as $q = 4\pi \sin(\theta)/\lambda$, with a scattering angle of 2$\theta$.

Differential scanning calorimetry (DSC) experiments (Figure S1) were carried out on a Netzsch DSC 200 F3 Maia apparatus with heating and cooling-rates of 1 K min$^{-1}$ under nitrogen atmosphere, using 20 mL aluminum pans with holes.

Table 1. Number average molar mass ($M_n$) of the polymers, degree of polymerization ($DP$), polymer contour length ($L_c$), polymer end-to-end distance ($R_e$), polymer volume fraction ($\phi_{polymer}$), hard-to-hard domain distance by SAXS ($d_{SAXS}$) and by AFM ($d_{AFM}$) for the three polyurea elastomers.

|  | ED-2000 | ED-4000 | ESD-2001 |
|---|---|---|---|
| $M_n$ (g mol$^{-1}$)[a] | 2056 | 4000 | 2050 |
| $DP$[a] | 34 | 69 | 34 |
| $L_c$ (nm)[b] | 12.0 | 24.4 | 12.0 |
| $R_e$ (nm)[c] | 8.9 | 18.1 | 8.9 |
| $\phi_{polymer}$[d] | 0.12 | 0.07 | 0.12 |
| $d_{SAXS}$ (nm) | 5.23 | 7.56 | --- |
| $d_{AFM}$ (nm) | 5.58 | 7.94 | --- |

[a] Values obtained by titration. [b] All-trans conformation of the fully stretched polymer. [c] estimated by simulating a *ca.* 5 repeating units helical structure. [d] Values from the stoichiometry used to prepare the samples.

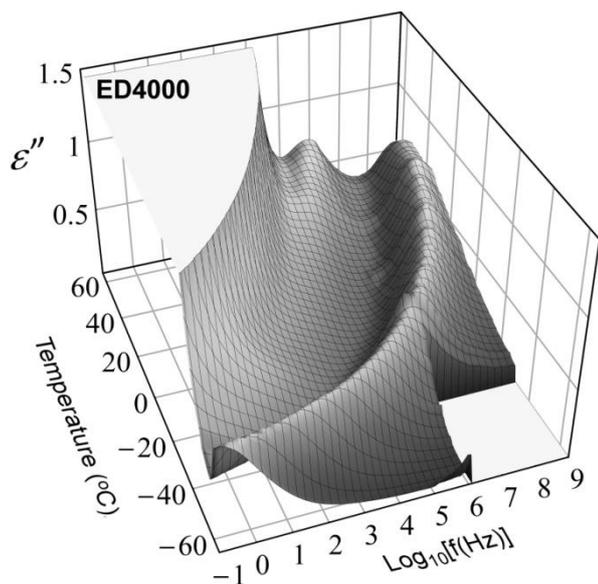

Figure 2. Three-dimensional plot of the dielectric loss vs. temperature and logarithm of the frequency for the ED-4000 sample.

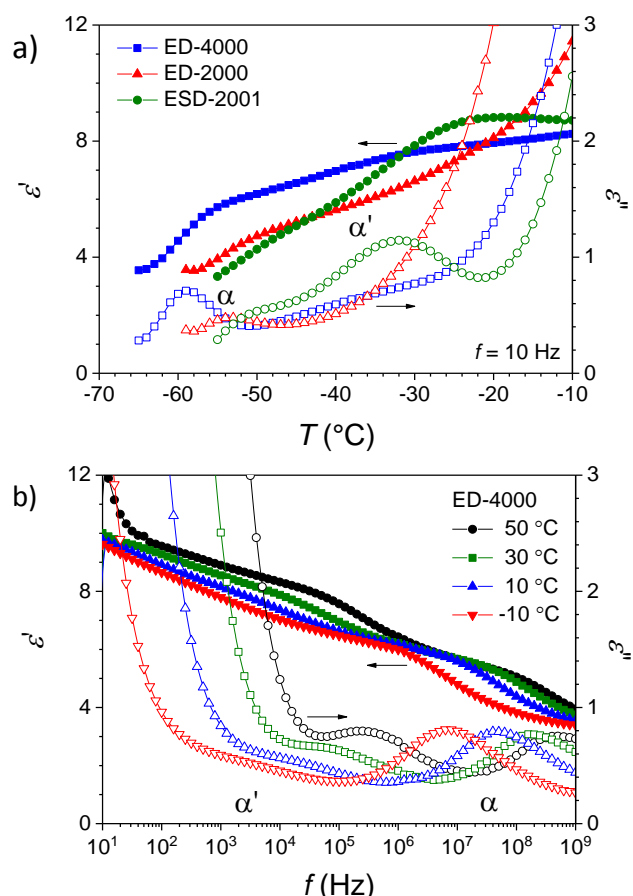

Figure 3. a) Temperature dependence of the real ($\varepsilon'$) and imaginary ($\varepsilon''$) components of the complex dielectric permittivity at 10 Hz, for the three elastomeric samples. b) Complex dielectric permittivity vs frequency in logarithmic scale for the ED-4000 sample at different temperatures. In both, full and empty symbols correspond to the real and imaginary component, respectively.

# 3 Results

All characteristics for the three elastomeric samples such as the number average molar mass ($M_n$) of the polymer, the corresponding degree of polymerization (*DP*), the polymer contour length ($L_c$), the polymer end-to-end distance ($R_e$), the polymer volume fraction ($\phi_{polymer}$) are reported in Table 1. Moreover, since polyurea elastomers are microphase separated materials with soft and hard domains, the hard-to-hard domain distance was evaluated by SAXS ($d_{SAXS}$) measurements[15] where values of 5.2 and 7.6 nm for the ED-2000 and ED-4000 elastomers were found respectively together with the amorphous nature of the polymer backbone as shown from the WAXS profiles (Figure S2). Complementary information was gained from AFM experiment on bulk or on casted elastomeric films (Figures 1, S3 and S4). AFM images of ED-2000 and ED-4000 show a nanoscale rod-like structure corresponding to the microsegregation of the hard domains from the soft matrix. The average hard-to-hard distance ($d_{AFM}$) was evaluated from the Fast Fourier Transform (FFT) of AFM images, and values of 5.6 and 7.9 nm for the ED-2000 and ED-4000 elastomers were obtained, respectively (Figure S5). On the other hand, the sample ESD-2001 showed neither any characteristic peak in SAXS nor in the AFM-FFT evaluation. The reason for the absence of such characteristic distance is due to the impossibility to establish hydrogen bonds between the urea motives because of the presence of bulky groups in the secondary amine of the polymer. Thus, the formation of hydrogen bonds in the ED-2000 and ED-4000 elastomeric samples should influence the mechanical properties when compared to the samples ESD-2001 without hydrogen bonds.[16]

Molecular dynamics in the ED-4000, ED-2000 and ESD-2001 polyurea elastomers were investigated by means of broadband dielectric spectroscopy over a wide temperature range. Frequency and temperature dependencies of the dielectric losses for the ED-4000 polyurea elastomer are shown in Figure 2, and the equivalent representations are given in Figure S6 for both ED-2000 and ESD-2001 polyurea elastomers. At first glance, all three polymer networks exhibit similar behaviour, *i.e.*, a dielectric spectrum characterized by two relaxation processes, which can be associated to the segmental chain dynamics i) of the less constrained chain segments embedded in the soft domains (relaxation at higher frequencies or $\alpha$) and ii) of the most restricted chain segments directly attached to the hard domains (relaxation at lower frequencies or $\alpha'$), as previously observed for polyurea systems by dielectric spectroscopy[17,18] and from dynamic mechanical analysis measurements.[19] However, clear differences are also evident when comparing the results obtained from the three samples as shown in the isochronal representation of Figure 3a. The $\alpha'$-relaxation is predominant in ESD-2001, while for ED-4000 and ED-2000 the $\alpha$-relaxation and the $\alpha'$-process appear as a pronounced and a broad peak, respectively. However, both relaxation processes can be clearly discerned in the isothermal representations in Figure 3b and S7. At temperatures above 40 °C, the shape of the peak related to the $\alpha'$-relaxation process becomes sharper, which can be explained by the progressive rupture of hydrogen-bonding between the urea motives, and thus by the increase of the dynamics of the chain segments directly attached to the urea motif.

For each temperature, the results were fitted to

$$\varepsilon(\omega) = \sum_k \Delta\varepsilon_k(\omega) + \varepsilon_\infty - i\frac{\sigma_0}{\omega\varepsilon_0} \qquad (1)$$

where $\varepsilon_\infty$ is the high frequency permittivity and $\sigma_0$ is the dc-conductivity. Both the $\alpha-$ and the $\alpha'$-relaxation process have been characterized by simultaneously fitting the real and imaginary dielectric permittivity to the Havriliak-Negami (HN) function:

$$\Delta\varepsilon_k(\omega) = \frac{\Delta\varepsilon_k}{\left[1 + (i\omega\tau_{HN})^{a_k}\right]^{b_k}} \qquad (2)$$

where $\Delta\varepsilon_k$ is the strength of each mode. The parameters $a_k$ and $b_k$ ($0 < a_k \leq 1$ and $0 < b_k \leq 1$) describe the broadness and symmetry of the relaxation spectra, respectively. The frequency of maximal loss is related to the shape parameters by the following equation:[20]

$$f_{\max,k} = \frac{1}{2\pi\tau_{HN,k}}\left[\sin\left(\frac{\pi a_k}{2+2b_k}\right)\right]^{1/a_k}\left[\sin\left(\frac{\pi b_k a_k}{2+2b_k}\right)\right]^{-1/a_k} \qquad (3)$$

It should be noted that in all three samples there is a great increase of both $\varepsilon'$ and $\varepsilon''$ at low frequencies, which arises not only from dc-conductivity but also can be related to the interfacial polarization at the internal boundaries between the hard and soft domains. This contribution masks the low frequency segmental relaxation and makes difficult to unambiguously determine the shape parameters for the low frequency relaxation process or $\alpha'$. To limit its impact in the results, a Debye relaxation could be easily added at low frequencies (Figure S8 and S9) for samples ED-2000 and ESD-2001. However, this approach could not be applied for ED-4000, and the value of the electronic conductivity was determined in the modulus representation, and then fixed in Equation 1 (Figure S10).[20,21] The shape parameter values $a_k$ and $b_k$ are summarized in Table 2 for the two relaxation processes and for the three samples.

The obtained relaxation frequencies of both relaxation processes as function of temperature for the three samples

Table 2. Relaxation shape parameters, $a_k$ and $b_k$, for both the $\alpha$ and $\alpha'$-processes, and for the three elastomeric samples.

|  |  | $a_k$ | $b_k$ |
|---|---|---|---|
| ED-4000 | $\alpha$ | 0.70-0.75 | 0.54-0.60 |
|  | $\alpha'$ | 0.40-0.50 | 0.6 |
| ED-2000 | $\alpha$ | 0.65-0.70 | 0.50-0.63 |
|  | $\alpha'$ | 0.35-0.70 | 0.6 |
| ESD-2001 | $\alpha$ | 0.40-0.65 | 0.65 |
|  | $\alpha'$ | 0.50-0.70 | 0.7 |

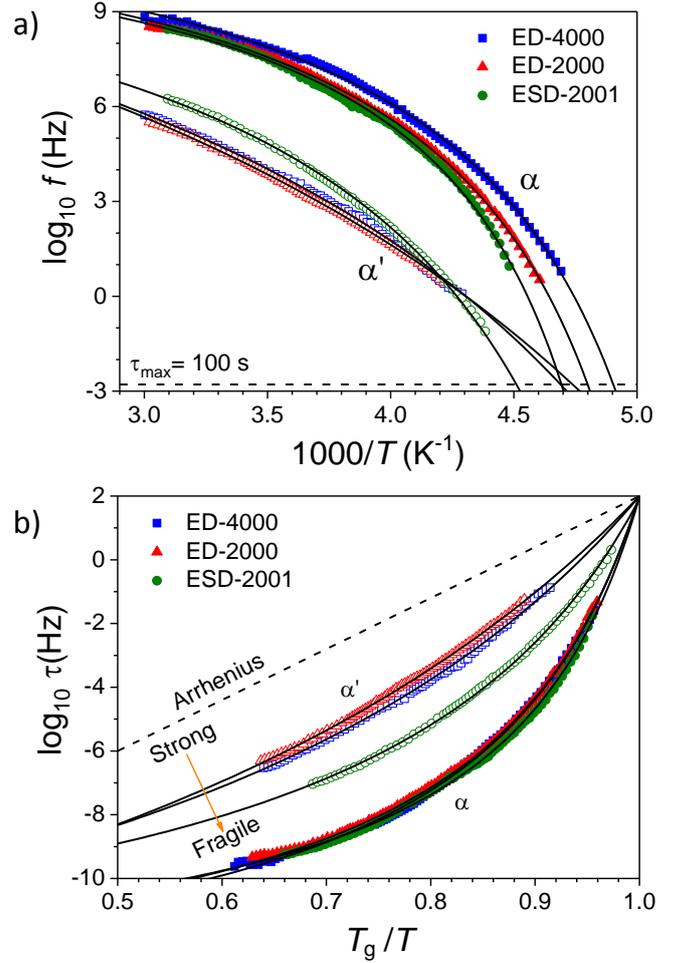

Figure 4. a) Arrhenius plot of the frequency of the maximum loss for the $\alpha$- (filled symbols) and the $\alpha'$-process (empty symbols) - as deduced from the fits to Eq. 1 of the three elastomeric samples, together with the corresponding fits to the VFT law. b) Angell's plot of the relaxation times for the $\alpha$- (filled symbols) and the $\alpha'$-process (empty symbols) for the three elastomeric samples, together with the corresponding fits to the VFT law.

are gathered in Figure 4a, which follow the Vogel-Fulcher-Tamman law (VFT)

$$f_k = f_{0,k}\exp\left\{-\frac{B_k}{T-T_{0,k}}\right\} \qquad (4)$$

where the pre-exponential factor $f_{0,k}$, the so-called Vogel temperature $T_{0,k}$, and the exponential factor $B_k$ (related to the fragility and to the apparent activation energy) are the fitting parameters. The dielectric glass transition temperature $T_g$ is calculated when the corresponding structural relaxation time is $\tau_{\max} = 100$ s. Although in calorimetric measurements a single glass transition is observed, dielectric measurements clearly show two segmental relaxation processes associated with a dielectric glass transition. The best VFT fits are plotted in Figure 4a, and the corresponding fitting parameters together with the obtained $T_g$ values are reported in Table 3, and compared to those obtained by dynamic mechanical analysis (DMA) and by differential scanning calorimetry (DSC, Figure S1) measurements.[13,15,16] It should be pointed out that, in DMA measurements, the storage and loss moduli are obtained at a given frequency as a function of temperature and the VFT behaviour is assumed in the fitting procedure. Thus, it is noteworthy that despite the difference between the methods, values are in reasonable agreement.

Table 3. The Vogel-Fulcher-Tamman (VFT) fitting parameter, $B$, $\tau_0$ and $T_0$, and the $T_g$ values obtained by of broadband dielectric spectroscopy, dynamic mechanical analysis (DMA) and differential scanning calorimetry (DSC).

|  | ED-4000 | | ED-2000 | | ESD-2001 | |
| --- | --- | --- | --- | --- | --- | --- |
|  | α | α' | α | α' | α | α' |
| $B$ (K) | 945±18 | 2477±87 | 944±12 | 3081±70 | 818±25 | 1423±39 |
| $\tau_0$ (s) | $10^{-12.4}$ | $10^{-12.0}$ | $10^{-12.2}$ | $10^{-12.7}$ | $10^{-11.9}$ | $10^{-11.1}$ |
| $T_0$ (K) | 175±1 | 137±2 | 180±1 | 120±2 | 187±1 | 175±1 |
| $T_g$ (K) | 204±1 | 213±7 | 208±1 | 211±5 | 213±1 | 222±3 |
| $T_{g,\text{DMA}}$ (K) | 201±1 | 222±1 | 199±1 | 208±2 |  |  |
| $T_{g,\text{DSC}}$ (K) | 206±1 | - - | 213±1 | - - | 214±1 | - - |
| $m$-index | 103±9 | 39±6 | 102±10 | 34±4 | 115±12 | 62±8 |

In Figure 4a, the α-relaxation process looks similar for the three samples at high temperatures, and the mobility is slightly but noticeably higher for the sample ED-4000 due to its higher segmental molecular weight. However, such difference is enhanced as the temperature decreases. On the other hand, the slow α'-relaxation process behaviour clearly differs between the samples with hydrogen bonding (ED-4000 and ED-2000) and the sample without (ESD-2001), which has a larger and more thermally activated characteristic frequency. This matches the picture of nano-segregated hard domains dispersed in a soft matrix, with the faster α-relaxation process related to the mobility of chain segments embedded in the soft domains and the slow α'-relaxation process arising from the restricted mobility of those chain segments in the vicinity of the hard domains. In this way, hydrogen-bonds further decrease the mobility of the restricted chain segments, thus the rupture of the physical interaction between them is expected to result in similar mobility for the three samples at temperatures higher than 70 °C. The longer segmental molecular weight in sample ED-4000 results in a higher mobility for the chain segments in the soft domain, corresponding to the α-relaxation process, as seen in Figure 4a. Thus, the mobility at high temperatures is similar for both the elastomeric samples ED-2000 and ESD-2001 with comparable segmental molecular weight, and increasingly differ on cooling due to the constrains imposed on the dynamical correlation length when hydrogen-bonding is formed for the sample ED-2000 which at the same time decreases $T_g$.

In order to quantify the change on the dynamics when approaching the $T_g$, we can make use of the fragility metric, which describes to which extent dynamics differ from the basic Arrhenius behavior. Following the formalism introduced by Angell,[22–26] the fragility index is defined by the slope value at $T_g$ for a $\log_{10} \tau$ vs. $T_g/T$ representation (Figure 4b). Minimum $m$ value is assigned to conventional Arrhenius behaviour $\tau = \tau_0 \exp\{E/k_B T\}$, which corresponds to $m = 16$ for the pre-factor $\tau_0 = 10^{-14}$ s. According to this metric, glass-forming materials can be classified in "fragile" ($m > 50$) and "strong" ($m < 30$). In the present case, all materials show two segmental relaxation processes, whose dynamics freeze at different $T_g$, and thus a fragility value $m$ can be obtained for each relaxation mode. Fits to VFT law (Figure 4b) allow to calculate the fragility index $m$ as:

$$m = \left[\frac{d(\log_{10} \tau_k)}{d(T_{g,k}/T)}\right]_{T=T_{g,k}} = \frac{B_k}{\ln(10) T_g (1 - T_0/T_g)^2} \quad (5)$$

The $m$ values obtained for ED-4000, ED-2000 and ESD-2001 are given in Table 3. For all three materials, values slightly above 100 for fast α-relaxation process were obtained. However, interesting differences can be noticed in the case of the slow α'-relaxation process. Fragility in this case, is greatly affected by the hydrogen-bonds between the urea motives, taking values around 35 for the ED-2000 and ED-4000, and almost double for EDS-2001, which accounts for the restricted dynamics of the chain segments near the hard domains.

## 4. Information from the shape parameters of the α-mode

The picture provided by the VFT analysis in the previous section can be refined if the relaxation frequency profiles are also taken into account. The conductivity component characterizing our systems, however, is too large to let the shape parameters of the α'-process be assigned without ambiguity. On the contrary, the α-process contribution is sufficiently well defined in order to allow for a reliable analysis of the cooperativity. Before doing this, we give a brief outline of the theoretical scheme that underlies our data treatment; further and more thorough details can be found in reference 7.

**Basic description of segmental relaxation**

The α dynamic regime in polymeric melts refers to a condition whereby the diffusional motion of segments is promoted by activated processes. The energy threshold ζ of a rearranging unit, however, is not constant in time, as it varies continuously due to local configurational changes; rather, each unit explores a distribution $p(\zeta)$ of such barriers. In a stationary state, a constant excess chemical potential $\overline{\Delta\mu}(T) > 0$ is then established, which describes the deviation of the liquid from equilibrium at the temperature $T$.

Configurational transitions take place by facilitation from precursory (or pre-transitional) excited states in which the rearranging units move about fixed sites, probing the $\zeta$-space that is locally accessible. Such non-diffusive dynamics is characterized by a certain energy per unit which has been gained by fluctuation; when this energy is so large that the probability of the excited state gets lower than that associated to a completely unconstrained mobility state - with ideal specific configurational entropy $s_c$ -, then a *small* number $z$ of units may acquire a diffusional component in their overall motion (in other words, the acquisition of a diffusional component in the motion is a spontaneous process). The local constraints that have contributed until then to hinder *large-scale* configurational changes in the surroundings deplete; more units start diffusing, further disentailing the neighboring ones in a sort of a cascade process. Facilitation, as just described, eventually leads to the formation of a high mobility spot - the Cooperatively Rearranging Region (CRR) - surrounded by glassy domains (possibly, local conformations characterized by high rearrangement barriers are pushed away from the spots' bodies to bank up at some liquid/glass interface). The number of units forming such mobility domains is in the order of

$$N_\alpha \approx z(\lambda + 1) \quad (6)$$

where $\lambda$ (in general $> z$) is the Lagrange multiplier associated to the condition that $\overline{\Delta\mu}$ is held constant when extremizing the thermodynamic potential for the derivation of $p(\zeta)$. More intuitively, the $z$ units may initiate a large-scale rearrangement provided other $\lambda$ units are dynamically correlated with each of them.

The nature of the CRRs is fluctuational; they form and are subsequently re-absorbed in the quasi-glassy liquid, so their total number is a non-conserved quantity. This means that a chemical potential can be associated to them and that it must be zero, implying on turn that the relation

$$\lambda \cong \left[1 - w(\zeta^*)\right] \frac{\zeta^*}{T s_c} \quad (7)$$

holds, where $w \equiv w(\zeta)$ is the probability that a unit has an energy not lower than $\zeta$, and $\zeta^*$ is the average barrier height that each unit of a CRR has to overcome in order that facilitation initiates, leading eventually to a large scale configurational change. Note that, in proximity of $T_g$, $\zeta$ is in fact close to the average energy per unit needed to initiate the large scale rearrangement from a non-diffusive pre-transitional dynamic state. This relation is an important indirect link between $\zeta^*$ and $N_\alpha$.

Concerning the CRR shape, the scheme proposed in reference 7 does not provide any prescription; thus, besides the common picture of compact structures,[3,4] the possibility of branched domains can also be envisaged.[27]

**Data analysis**

Information about $\lambda$ and $z$ is carried by the relaxation function $\varphi_{exp}(t)$ associated to the α-process; $\varphi_{exp}(t)$ can be obtained from the Fourier cosine transform of the HN frequency loss profile in the complex permittivity $\varepsilon(\omega)$. The obtained expression can be then fitted with the function

$$\varphi_{th}(t) = \int_0^\infty d\zeta\, p(\zeta) \exp\left\{-t\nu^* e^{-z\Delta\mu(\zeta)/k_B T}\right\} \quad (8)$$

where $\nu^*$ is an attempt rate and $\Delta\mu(\zeta) = -k_B T \ln w(\zeta)$ is the excess chemical potential associated to a barrier height $\zeta$ (its $p$-weighted average yields the chemical potential $\overline{\Delta\mu}$ mentioned at the beginning of the preceding subsection). Of course, $\lambda$ enters $p(\zeta)$ as a parameter and in fact, not too far from $T_g$, $p(\zeta) \approx \exp\{-\lambda \Delta\mu(\zeta)/k_B T\}$.

The information provided by the analysis of $\varphi_{exp}(t)$ is related directly to the precursory state, since the parameters $z$ and $\lambda$ characterize the local dynamics immediately preceding the large scale rearrangements.[7] The attempt rate in Eq. 8 is expected to depend on $T$. For the present purposes, it suffices to treat it as a fitting parameter.

**Results and discussion**

Figure 5 shows that in all samples the rearrangement threshold eventually increases when the proper $T_g$ is approached on cooling. The abscissas report the difference $T - T_g$, and steeper growths are found where configurational restrictions are weaker; this means that rising $\zeta^*$ by enhancing the constraints is not equivalent to cooling a less constrained system, though it may appear so on qualitative grounds. If on the contrary this were the case, the $\zeta^*(T)$ functions of, *e.g.*, ED-2000 and ED-4000 could be made exactly overlapping by a suitable temperature shift, at least in the low-$T$ region (cf. the straight lines in Figure 5). As will be shown below, the same argument applies based on the cooperativity reported in Figure 6.

A $\zeta^*$ increase on cooling is in general expected, particularly in proximity of $T_g$, because a fraction of higher and higher local potential barriers is probed by the rearranging units.[7] On the other hand, the tendency to level off and upwards bend on heating cannot at present be self-consistently interpreted within our theory, and we can only guess for reasonable underlying mechanisms. We shall touch upon this point below.

Comparison of ED-2000 with ED-4000 shows that the energy barrier $\zeta^*$ at fixed $T - T_g$ is larger for shorter soft chains. Concurrently, both the central relaxation time and the peak width (in the −50 °C ≤ $T$ ≤ 30 °C interval) increase upon shortening the soft chain length. This conforms to constraining conditions in adsorption regime of nano-pore confinement,[28] that is, when the relaxing regions are strongly coupled with the walls of the confining volume; indeed, this coupling is represented here by the soft chain pinning to the hard domains.

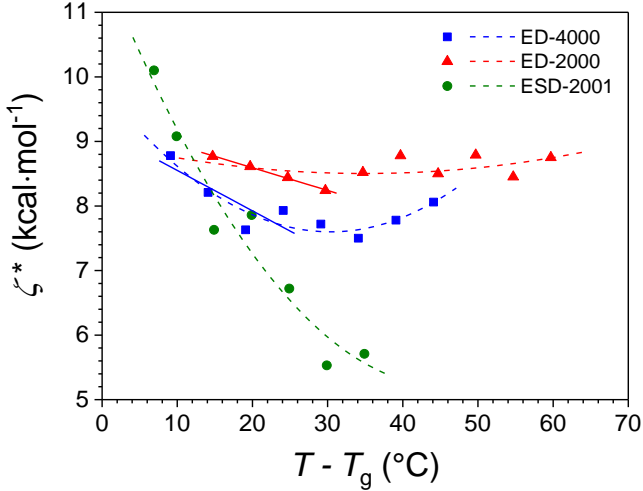
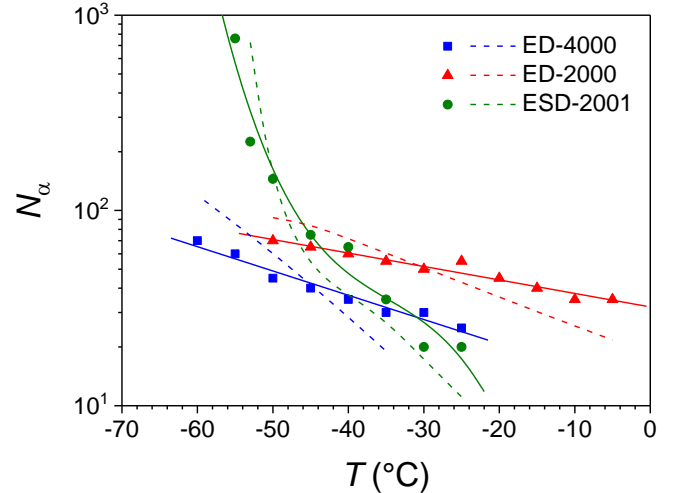

Figure 5. Average height of the rearrangement energy barrier as a function of $T–T_g$ (cf. Table 3 for the appropriate $T_g$ values). The 2nd order polynomial fits to the data are guides for the eye. The straight lines are linear fits to the four lowest temperature data points of ED-2000 and ED-4000.

Figure 6. Cooperativities as functions of the temperature after Eq. 6. Solid lines are fits, either linear or 3rd order polynomial, to the data (symbols) worked out after adjusting Eq. 8 to $\varphi_{exp}(t)$; they are just guides for the eye. The dashed lines are obtained by means of the data analysis adopted in refs 13 and 29, i.e., after Eq. 9 of the text.

A possible reason for the increase of $\zeta^*$ is the emergence - or increasing significance - of additional mechanisms, not directly related with facilitation, through which the energy of the pre-transitional state can return to the heat bath. In this case, the characteristic time facilitation takes to start from the excited state would compete with the lifetime of the precursor state itself, which is shortened by the presence of these additional paths of energy loss. Larger initial energies of the pre-transitional dynamic state - and also cooperativities, after Eq. 6 and 7 - would thus be needed in order that the energy of the precursor state still remains high enough to render facilitation thermodynamically advantageous (i.e., spontaneous) should it happen to initiate. Incidentally, the emergence of new paths of energy regression could be hypothesized also in the tentative explanation of the $\zeta$ increase at high temperatures noticed above, that is found on heating above temperatures specific of each of the samples. Of course, a reliable description of the heat transfer among different degrees of freedom (i.e., vibrational and configurational) would be required to support this hypothesis, but we must defer this issue to future work.

The overall slowing down of the ED-2000 dynamics compared to ED-4000 is, of course, associated with a shortening of the soft chains, since the way constraints are introduced and the nature of the soft/hard interface are the same for both. Such a situation seems to be counterintuitive, because a restriction of the available volume would be expected to also limit the mean size of the CRRs and, together with it, the activation energy for the configurational changes. On the other hand, cooperativity is not just a consequence of a "drag" effect characterizing the diffusional motion of rearranging units. Rather, what we observe is in fact the precursor dynamic state; the number of correlated units necessary to attract energy enough by fluctuation (via a sufficient number of available states, that is, a sufficient entropy) may well increase to contrast the effect of the constraints. The cooperativity estimates of Figure 6 show that indeed this may be the case.

For a comparison, we also report in Figure 6 the results found by means of the 4-point correlation function approach.[5,13,27,29] In that framework, indeed, a lower limit to the number of correlated units is estimated by the equation

$$N_{corr} = \frac{k_B}{\Delta c_p}\left[\frac{\beta(T)}{e}\right]^2\left[\frac{d\ln\tau_\alpha}{d\ln T}\right]^2, \qquad (9)$$

where $e$ is the Euler number, while $\beta$ and $\tau_\alpha$ parametrize the Kohlrausch-Williams-Watts (KWW) function, $\exp(-[t/\tau_{\alpha(T)}]^\beta)$, that reproduces our $\varphi_{exp}(t)$ at each temperature in the relevant time interval. The agreement between $N_\alpha$ and $N_{corr}$ has to be considered good, also from a quantitative viewpoint.

The cooperativity in ED-2000 increases on cooling at a lower rate than in ED-4000. In this respect, the behaviour shown in Figure 6 does not differ qualitatively from that of Figure 5, - except in that no upwards bending is found on heating at high temperatures. Considering the general tendency of cooperativity to grow on cooling, this difference in slope may be intuitively interpreted as a saturation effect of geometric nature, since the CRR growth is contrasted by the presence of chain pinning points and the soft domains in ED-2000 are smaller. Unfortunately, $\alpha$-relaxation patterns at temperatures closer to $T_g$ cannot be extracted in the present conditions, and a crossing between the ED-2000 and ED-4000 lines can only be guessed.

Focusing now on ESD-2001, Figure 6 shows that its cooperativity line crosses that of ED-2000 at $T \cong T_{cr} \equiv -40$ °C. The length of the soft chains is the same in both systems, but in ESD-2001 their ends are not pinned to fixed positions; rather, some limited mobility - though somewhat contrasted by the crosslinking units and the other chains they drag - is allowed. At $T > T_{cr}$, the cooperativity in the less constrained system is smaller, conforming to what has been found above comparing ED-2000 with ED-4000. On the other hand, a dichotomic behaviour emerges at $T < T_{cr}$: upon releasing configurational constraints with respect to ED-2000, we find that $N_\alpha$ decreases in ED-4000 and increases in ESD-2001. Clearly, qualitative arguments simply based on just the effectiveness of the constraints cannot help and we need to look at some other quantity for some discussion.

| 8

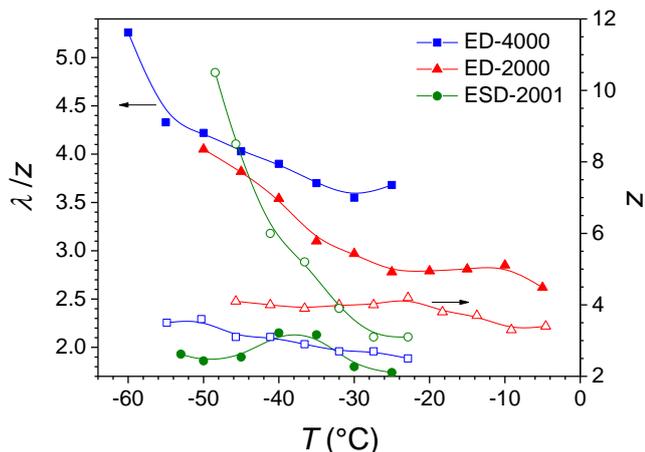

Figure 7. Ratio λ/$z$ (filled symbols) and $z$ (empty symbols) as functions of temperature obtained by fitting $\phi_{exp}$ with Eq. 8.

For the systems analyzed in a previous paper,[3] it has been noticed that in the presence of confinement the ratio λ/$z$ tends to lower. Figure 7 reports this quantity in all cases considered and shows that indeed ED-4000 and ED-2000, in which the relaxing chains have the same topology, follow this rule. In ESD-2001 instead (where the topology of the rearranging chains is different), λ/$z$ is always found to be less than in the others, changing just a little over the explored $T$-interval (filled circles); on the other hand, $z$ increases dramatically on cooling (open circles). At $T \cong T_{cr}$, a relatively smaller number of units ($z$) initiate the facilitation process in ED-2000, but each of them needs a comparatively large number of correlated units (λ) to be able explore the ζ-space before the large scale transition starts; the opposite is true in ESD-2001. Thus, although at this temperature the cooperativities are almost the same, the stronger mechanical constraints of ED-2000 tend to restrict the size ($z$) of the spots where the configurational transitions may initiate. Note that in these circumstances, the mean barrier height is also larger (cf. Figure 5).

The effect of the rigid crosslink units in ESD-2001 is twofold: on the one hand, they contribute to hinder the *configurational* degrees of freedom of the soft chains, though not as much as in ED-2000, and on the other hand, they couple mechanically all soft chains attached. Since our analysis probes the precursor state, we have direct evidence of a mechanical coupling in the absence of significant diffusing motion. When $T$ > $T_{cr}$, the precursory state involves regions of so limited an extension that the pinned ends of the soft chains are only marginally involved. In these conditions, ESD-2001 really appears to be less constrained compared to ED-2000, and the cooperativity turns out to be lower, approaching that of ED-4000.

## 5. Concluding remarks

The arrest of crystallization is probably the most elementary and widely known example of how structure may affect the primary relaxation dynamics in polymeric systems.[30] In that case, for semi-crystalline materials as well as confined systems, *e.g.*, crosslinked networks, elastomeric nano-composites, filled nano-porous materials, after the crystalline structure has evolved in a more or less controlled manner, the amorphous domains invariably reach a state in which the configurational degrees of freedom are frozen. In the systems considered here instead, polyurea materials with segregated hard domains form a continuous amorphous matrix in which the interface could influence the dynamics of the low glass transition domains; the configurational constraints are established at the outset and do not necessarily force the amorphous regions to vitrify.

Appropriate experimental methods, together with an exhaustive relaxation analysis, allowed for a detailed description of the interplay between the structure and the fast contribution to the primary relaxation of the amorphous domains. Concerning instead the slow dynamics of the semi-rigid layers surrounding the hard domains, one has to resort to different probes for the reasons illustrated previously. This may help to extract, by direct analysis in terms of both, the cooperativity and the rearrangement threshold energy ζ*, valuable information concerning, *e.g.*, the thickness $d_{\alpha'}$ of these interfacial layers. Indeed, the information extracted by the relaxation analysis adopted in the present case, depends only weakly on the probe, as it was shown in refs.[7,8]; this is supported also by the good agreement between the ED-2000 cooperativities worked out here and in ref.[13] by dynamic mechanical spectroscopy.

A reasonable upper bound to $d_{\alpha'}$ can be estimated in the case of the ED-2000 sample on the basis of the fast mode cooperativity (Tables S1, which considers low-$T$ extrapolated cooperativities, and S2, which refers to the actual data of Fig. 6). With reference to Fig. 1c, simulations yield a diameter of the hard core (blue colored circles in the figure) *ca.* 2.1 nm in the all-*trans* conformation of the hexamethylene chains from the crosslinker moiety. On the other hand, SAXS and Fourier transform AFM analyses yield a 5.2 to 5.6 nm, respectively, typical distance between the centers of these hard domains. Assuming that the cooperativity of the fast mode is close to saturation at the lowest probed temperature, one finds a diameter of about 2.5 nm for an approximately spherical CRR. This implies a surrounding layer of restricted mobility of thickness $0.4 < d_{\alpha'} < 0.5$ nm, which is compatible with the estimate of about 0.2 nm derived by independent considerations.[13] Such a value ($0.4 < d_{\alpha'} < 0.5$ nm) would correspond to at most two monomers in *trans* conformation. The possibility of probing the dielectric response at temperatures closer to the fast mode $T_g$, would made possible to work out a better estimate. However, the broadness of the relaxations and the disruption introduced by low frequency spurious contributions hinders the obtaining of reliable adjustments at those temperatures. The adoption of the same argument for the ED-4000 sample and using the lowest temperature available data or by extrapolating the values to temperatures just above the glass transition temperature (Figure 6) would clearly provide too a large upper bound because the cooperativity at that temperature is determined mainly by energetic exchange mechanisms, as if the relaxing domain were bulk. Stated otherwise, if one considers similar cooperativity values - say $N_\alpha \sim 80$ - in both ED-2000 and ED-4000, and the same thickness of the stiff layer $d_{\alpha'}$ is assumed for both systems due to the similarities in structure, it is found that a relatively large soft gap remains between a CRR and the surrounding hard domains in ED-4000; that is, saturation is not reached yet.

To conclude we point out that, although based on sound statistical mechanical arguments, the model adopted for the discussion of the results is merely descriptive, though it provides more details than the customary approaches. For example, recent application to the relaxation analysis of plastic crystals or "orientationally disordered crystals" (in which molecules' centers of mass are arranged in regular lattices, but the molecular directors are not) gave the possibility to discuss the effect of hydrogen bonding on cooperativity.[31] When dealing with confinement, in particular, geometric considerations alone cannot in general account for all the aspects of the process. In particular, as touched upon in the discussion on fast relaxation, the role of the threshold energy $\zeta^*$ seems to be important, and indeed a step further in the understanding of the cooperative dynamics would also need a satisfactory description of heat transfer among the different degrees of freedom. How efficient is the transfer of the pre-transitional energy towards the configurational modes, compared to other regression paths, is indeed central for a more complete understanding of confinement problems in general, just leaving aside the still lively problem of the super-arrhenian behavior of fragile liquids.

## Conflicts of interest

There are no conflicts to declare.

## Acknowledgements

This work was supported by the COST Action MP0902 (COINAPO – Composites of Inorganic Nanotubes and Polymers). N.S. thanks the "EU Horizon 2020 Framework Programme for Research and Innovation" for its support through the Marie Curie Individual fellowship No. 701558 (MagNem) at the Jozef Stefan Institute. Authors are especially grateful to Prof. M.R de la Fuente for providing access to the dielectric laboratory of the Universidad del País Vasco (UPV/EHU) and for stimulating discussions.

# Electronic Supporting Information

# Interplay between structure and relaxation in polyurea networks: the point of view from a novel method of cooperativity analysis of dielectric response

Nerea Sebastián,[a,b] Christophe Contal,[c] Antoni Sánchez-Ferrer,[*,d] Marco Pieruccini[*,e]

[a] *Jožef Stefan Institute, Department of Complex Matter - F7, Jamova cesta 39, SI-1000 Ljubljana, Slovenia.*

[b] *University of the Basque Country, Department of Applied Physics II, Apdo. 644, E-48080 Bilbao, Spain*

[c] *Institut Charles Sadron, Physics-Mechanics and Tribology of Polymers, 23 rue du Loess, BP 84047, F-67034 Strasbourg cedex, France.*

[d] *Swiss Federal Institute of Technology, Department of Health Sciences and Technology, IFNH, Schmelzbergstrasse 9, LFO, E29*
*CH-8092 Zürich, Switzerland.*

[e] *CNR, Istituto Nanoscienze, v. Campi 213/A, I-41125 Modena, Italy*

Corresponding author:

Marco Pieruccini (marco.pieruccini@nano.cnr.it)

Antoni Sánchez-Ferrer (antoni.sanchez@hest.ethz.ch)

**Table of Contents**

**Sample Characterization**

**Broadband dielectric spectroscopy**

**Procedure for determination of the dc conductivity contribution from the electric modulus.**

**Geometrical estimations**

**Table S2:** Cooperativity values ($N_\alpha$), radius of the cooperatively rearranging regions ($R_{CRR}$), soft domain average distance ($d_\alpha + d_{\alpha'}$), and thickness of the restricted mobility layer ($d_{\alpha'}$) at -50 °C for the samples ED-2000 and ED-4000.

**Sample Characterization**

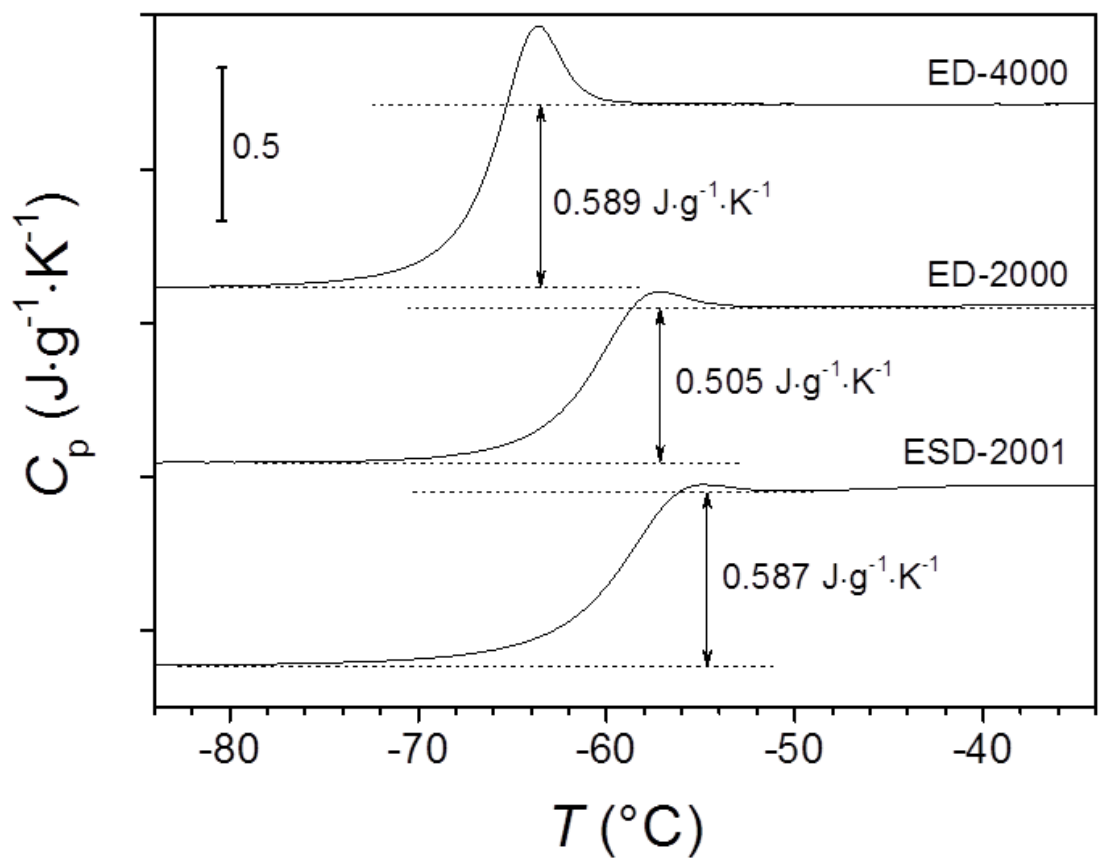

Figure S1. Heat capacity measurements of the three polyurea elastomers around glass transition on heating.

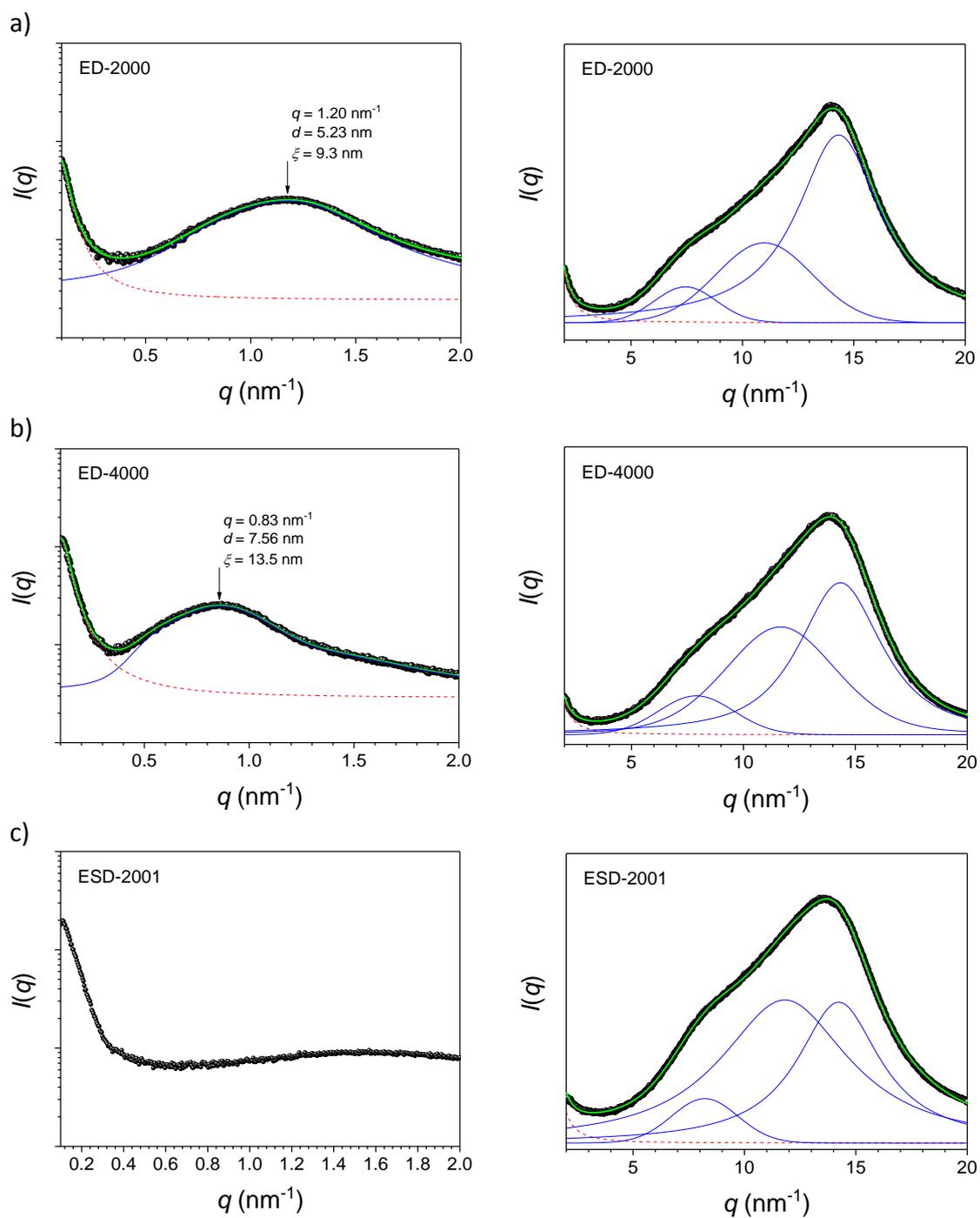

Figure S2: SAXS (left) and WAXS (right) intensity profile for of the three elastomeric samples a) ED-2000, b) ED-4000, and c) ESD-2001.

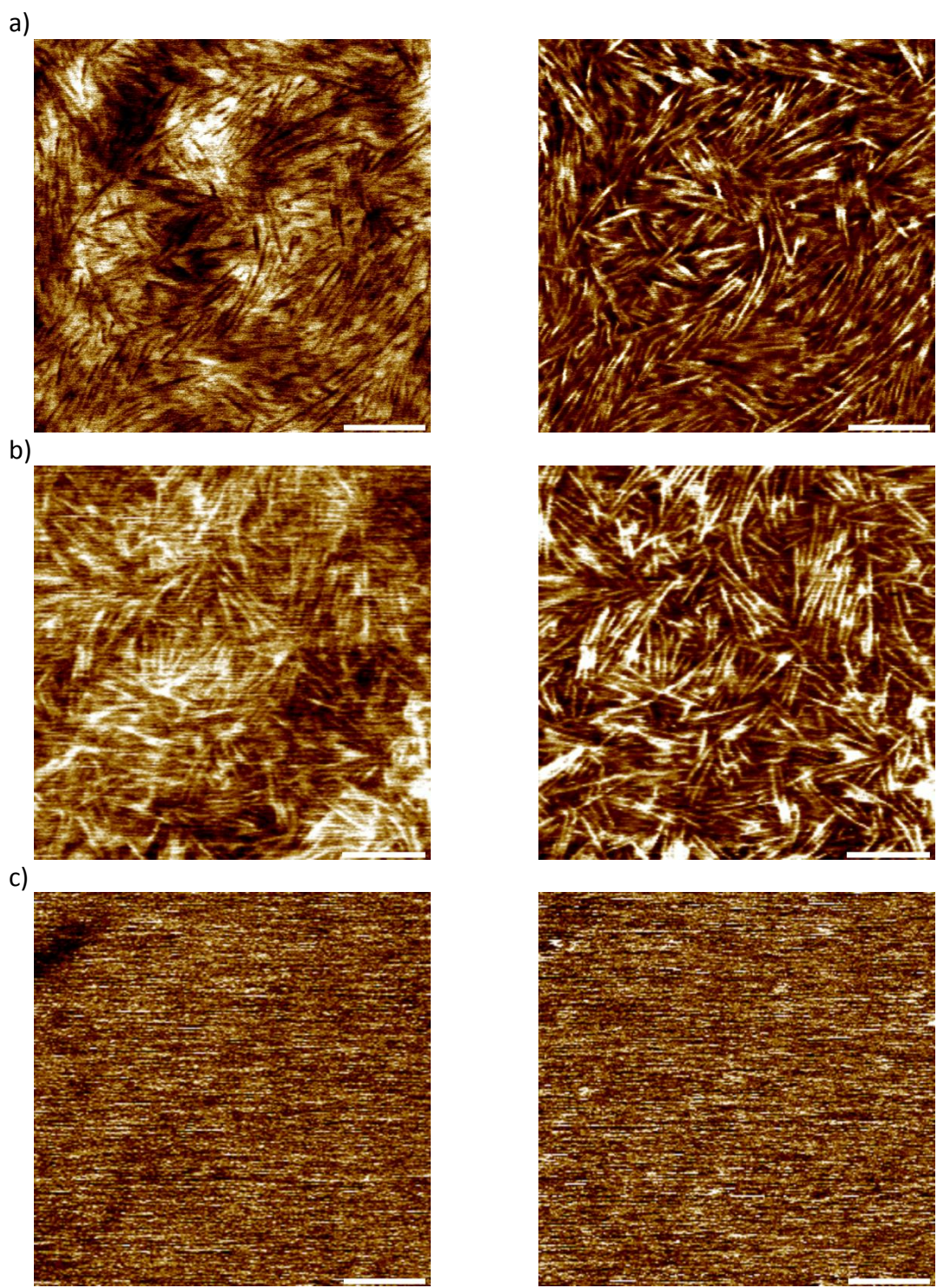

Figure S3. AFM height (left) and phase (right) profile image of the three bulk elastomeric samples a) ED-2000, b) ED-4000, and c) ESD-2001. *Note*: scale bar is 100 nm.

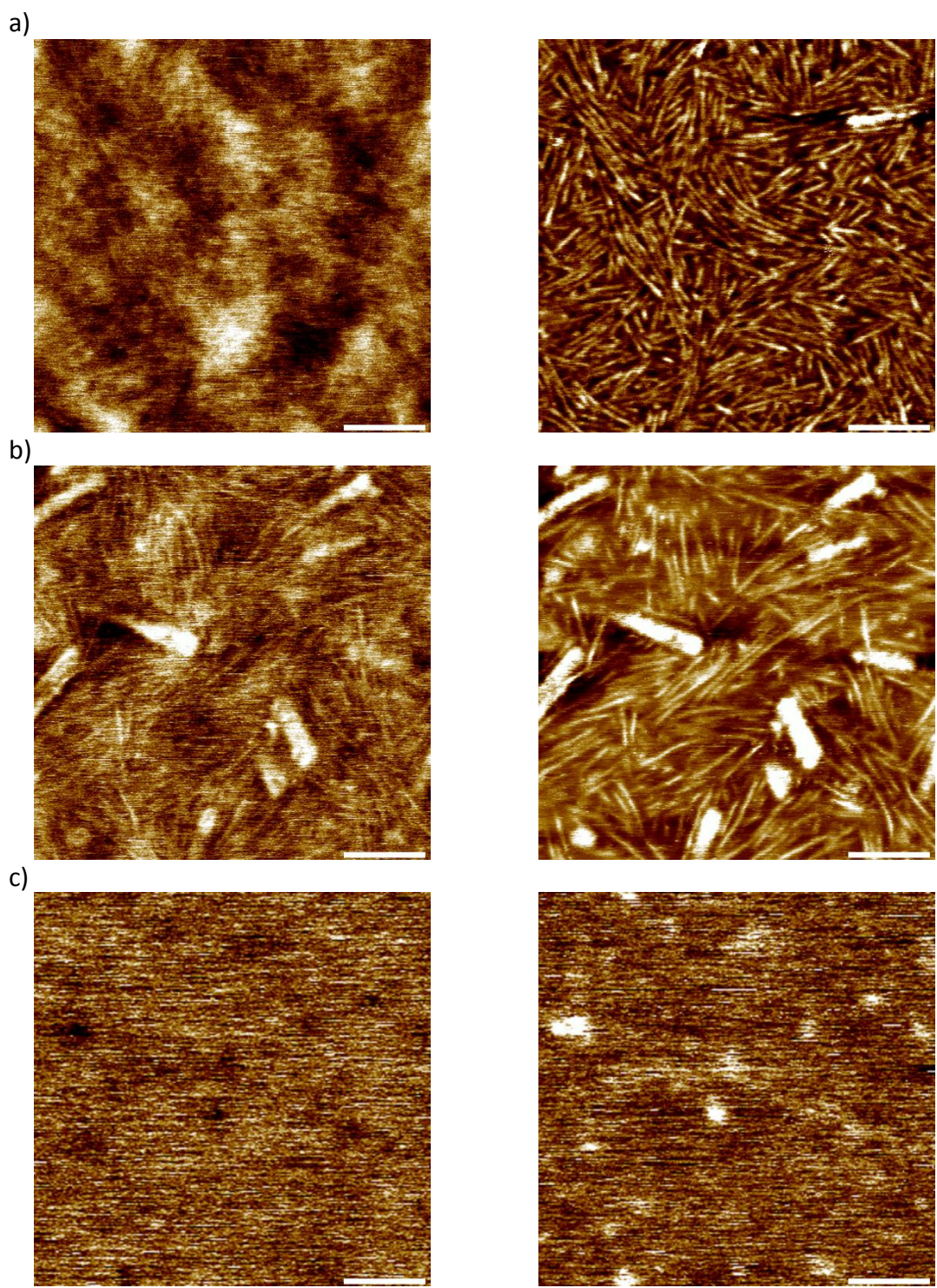

Figure S4. AFM height (left) and phase (right) profile image of the three casted elastomeric samples a) ED-2000, b) ED-4000, and c) ESD-2001. *Note*: scale bar is 100 nm.

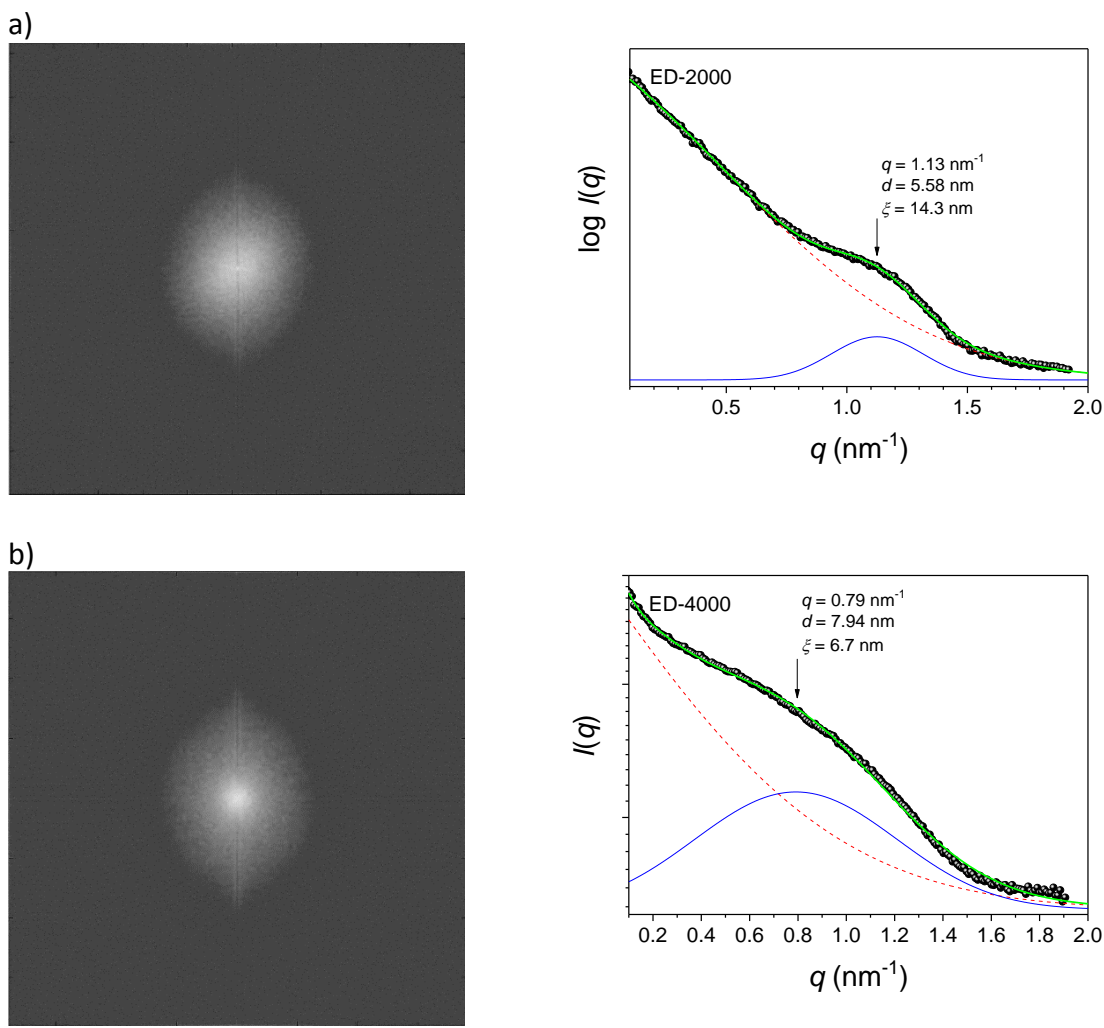

Figure S5. 2D (left) and 1D FFT plot (right) of the two elastomeric samples a) ED-2000, and b) ED-4000.

**Broadband dielectric spectroscopy**

a)
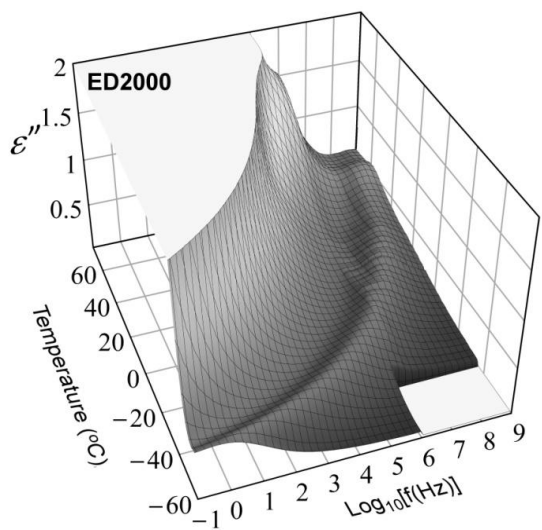

b)
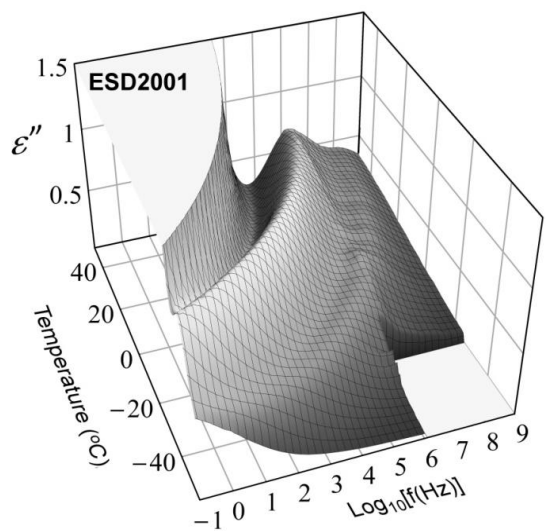

Figure S6. Three-dimensional plot of the dielectric losses vs temperature and logarithm of the frequency for a) ED-2000 and b) ESD-2001.

a)
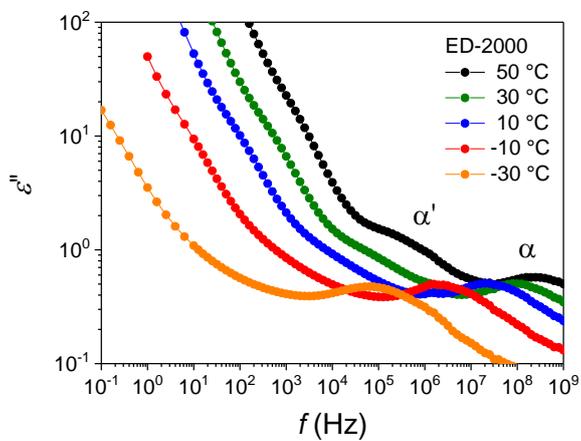

b)
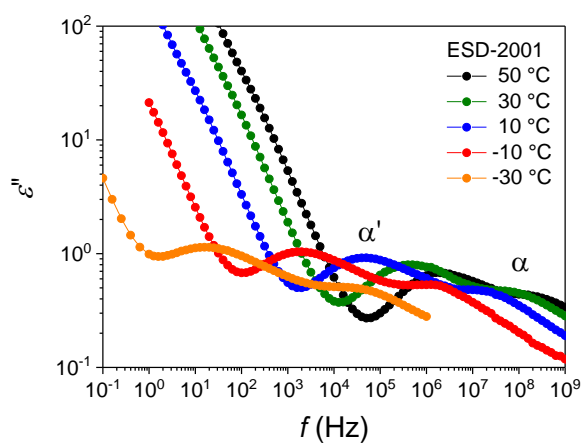

Figure S7. Dielectric losses vs logarithm of the frequency for a) ED-2000 and b) ESD-2001 at different temperatures.

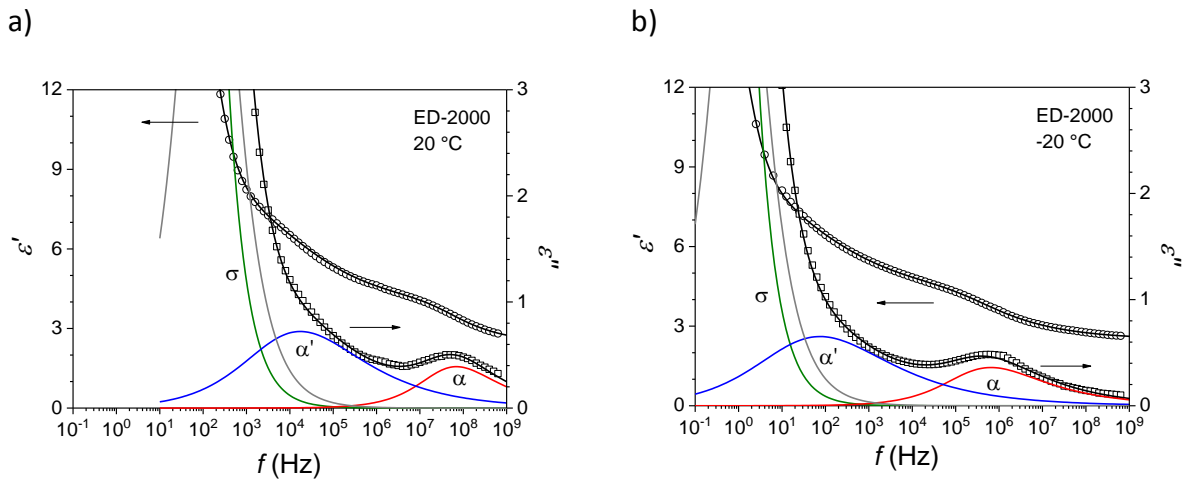

Figure S8. Real part (empty dots) and imaginary part (empty squares) of the dielectric permittivity versus frequency for ED-2000 at a) 20 °C and b) -20 °C. Solid black lines represent best fit to equation 1 (main text), and the color lines the deconvolution into the elementary modes α and α'. Green lines correspond to the dc conductivity contribution. Low frequency mode corresponds to MWS polarization.

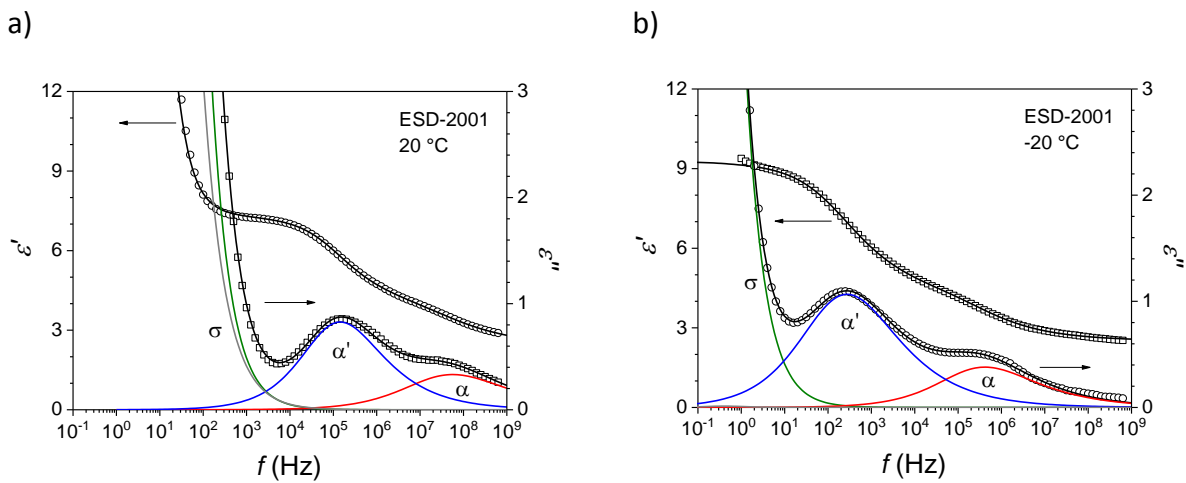

Figure S9. Real part (empty dots) and imaginary part (empty squares) of the dielectric permittivity versus frequency for ESD-2001 at a) 20 °C and b) -20 °C. Solid black lines represent best fit to equation 1 (main text), and the color lines the deconvolution into the elementary modes $\alpha$ and $\alpha'$. Green lines correspond to the dc conductivity contribution. Low frequency mode corresponds to MWS polarization.

**Procedure for determination of the dc conductivity contribution from the electric modulus.**

The complex electric modulus is defined as the inverse of the complex dielectric permittivity:

$$M^*(\omega) = 1/\varepsilon^*(\omega) = M'(\omega) + iM''(\omega) \qquad (E1)$$

which gives

$$M'(\omega) = \frac{\varepsilon'(\omega)}{\varepsilon'^2(\omega) + \varepsilon''^2(\omega)}$$
$$M''(\omega) = \frac{\varepsilon''(\omega)}{\varepsilon'^2(\omega) + \varepsilon''^2(\omega)} \qquad (E2)$$

For pure dc-conductivity, there is no electronic conduction contribution to $\varepsilon'(\omega)$, while $\varepsilon''(\omega) = \sigma_0/\omega\varepsilon_0$. It can be demonstrated that

$$M''(\omega) = M_\infty \frac{\omega\tau_{cond}}{1+(\omega\tau_{cond})^2} \qquad (E3)$$

where $\tau_{cond} = \varepsilon_0\varepsilon_\infty/\sigma_0$. Equation E3 is comparable to the imaginary component of a Debye relaxation process.

The electric modulus can be fitted then to

$$M^*(\omega) = M_\infty + \frac{\Delta M_{cond}}{1+(i\omega\tau_{cond})} + \sum_k \frac{\Delta M_k}{\left(1+(i\omega\tau_{M,k})^{a_{M,k}}\right)^{b_{M,k}}} \qquad (E4)$$

where $\tau_{M,k}$ are the characteristic relaxation times in the modulus representation and $\Delta M_k$ are the amplitudes for the electric modulus given by the difference between the limiting low and high frequency values of each process. Parameters $a_{M,k}$ and $b_{M,k}$ stand for the symmetric and asymmetric broadening of the relaxation. It should be pointed out, that the shape parameters in E4 do not correspond to those of the permittivity representation in eq. 1 of the main text. By fitting the electric modulus to expression E4, $\sigma_0$ can be determined as $\varepsilon_0\varepsilon_\infty/\tau_{cond}$. This value can then be fixed when performing the independent analysis of the complex dielectric permittivity according to equation 1 in main text. Examples of this approach are given in Figure S10.

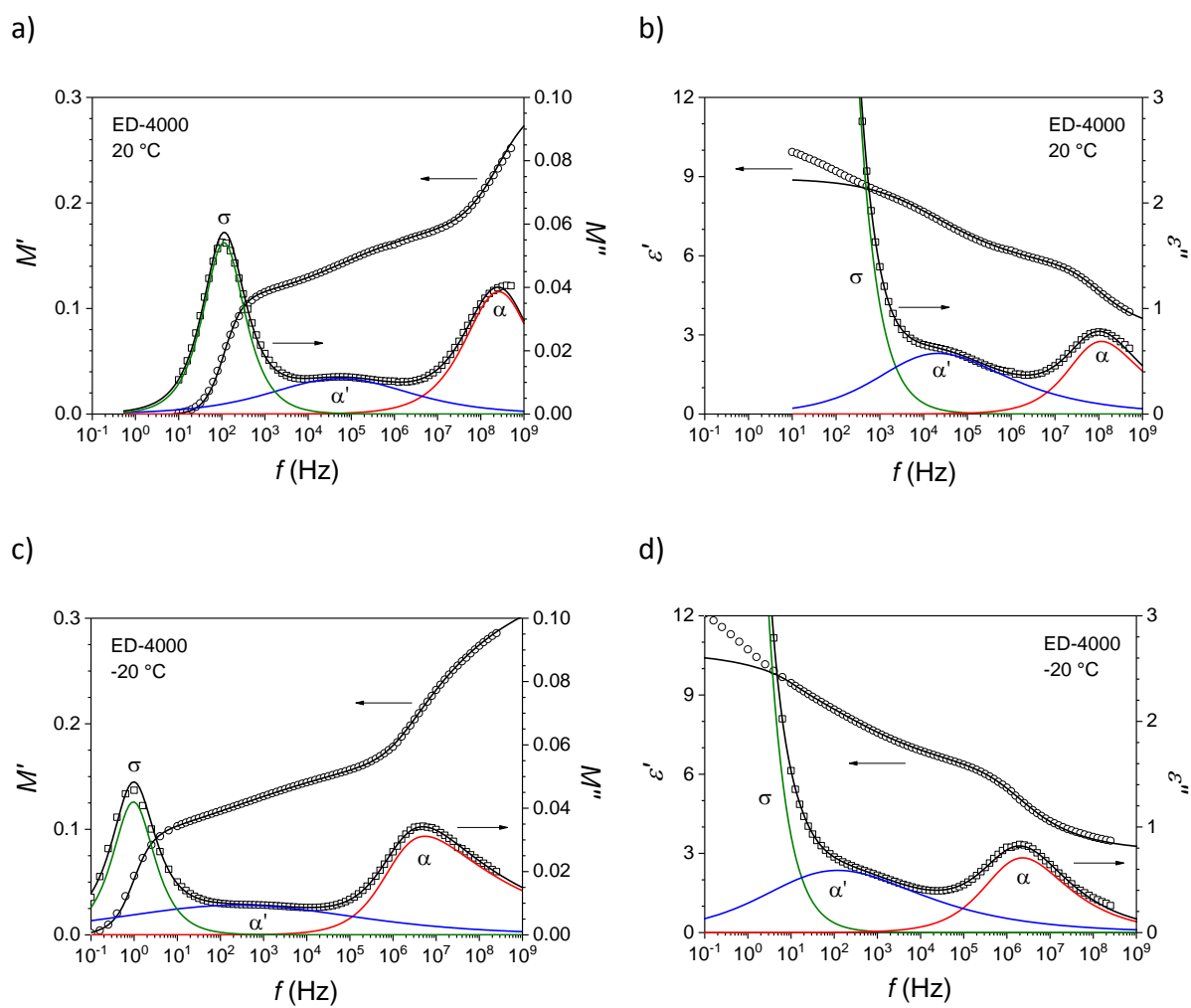

Figure S10. Real part (empty dots) and imaginary part (empty squares) of the electric modulus (a and c) and of the dielectric permittivity (b and d) versus frequency for ED-4000 at a) and b) 20 °C and c and d) -20 °C. In the Modulus plots solid black lines represent best fit to Equation E4 (main text) and while in the permittivity plot they correspond to the results of independently fitting to equation 1 in the main text. Red and blue lines correspond to the devoncolution into the elementary modes $\alpha$ and $\alpha'$ obtained at each independent fit. Green lines correspond to the dc conductivity contribution.

Table S1. Cooperativity values ($N_\alpha$), radius of the cooperatively rearranging regions ($R_{CRR}$), soft domain average distance ($d_\alpha + d_{\alpha'}$), and thickness of the restricted mobility layer ($d_{\alpha'}$) at the temperature just above the glass transition temperature for the samples ED-2000 and ED-4000.

| Sample | $N_\alpha$ | $T$ (°C) | $R_{CRR}$ (nm) | SAXS | | AFM | |
|---|---|---|---|---|---|---|---|
| | | | | $d_\alpha + d_{\alpha'}$ (nm) | $d_{\alpha'}$ (nm) | $d_\alpha + d_{\alpha'}$ (nm) | $d_{\alpha'}$ (nm) |
| ED-2000 | 84 | -60 | 1.25 | 3.32 | 0.41 | 3.54 | 0.52 |
| ED-4000 | 80 | -67 | 1.23 | 5.52 | 1.53 | 5.82 | 1.68 |

Table S2. Cooperativity values ($N_\alpha$), radius of the cooperatively rearranging regions ($R_{CRR}$), soft domain average distance ($d_\alpha + d_{\alpha'}$), and thickness of the restricted mobility layer ($d_{\alpha'}$) at -50 °C for the samples ED-2000 and ED-4000.

| Sample | $N_\alpha$ | $T$ (°C) | $R_{CRR}$ (nm) | SAXS | | AFM | |
|---|---|---|---|---|---|---|---|
| | | | | $d_\alpha + d_{\alpha'}$ (nm) | $d_{\alpha'}$ (nm) | $d_\alpha + d_{\alpha'}$ (nm) | $d_{\alpha'}$ (nm) |
| ED-2000 | 71 | -50 | 1.18 | 3.32 | 0.48 | 3.54 | 0.59 |
| ED-4000 | 49 | -50 | 1.05 | 5.52 | 1.71 | 5.82 | 1.86 |